\magnification\magstep1
\font\largebf=cmbx12 scaled \magstep1
\font\bigbf=cmbx12
\font\twelverm=cmr12
\footline={\hfil}
\headline={\ifodd\the\pageno \hfil\the\pageno
           \else \the\pageno\hfil
           \fi}
\openup2\jot

\centerline{\largebf Loop Amplitudes in Supergravity by Canonical Quantization}
\bigskip
\centerline{P.D. D'Eath}
\medskip                
\centerline{Department of Applied Mathematics \& Theoretical Physics}
\smallskip
\centerline{Silver Street}
\centerline{Cambridge CB3 9EW}
\centerline{UK}
\bigskip\bigskip
\centerline{Talk given at the 2nd INTAS meeting, Observatoire de Paris}
\centerline{1-3 May,1998, to be published in the Proceedings, ed. N.Sanchez}   \bigskip\bigskip        

\centerline{\bigbf ABSTRACT}
\smallskip\noindent
  Dirac's approach to the canonical quantization of
constrained systems is applied to  $N = 1$ supergravity, with or
without gauged supermatter.  Two alternative types of boundary
condition applicable to quantum field theory or quantum gravity are
contrasted. The first is the `coordinate' boundary condition as used
in quantum cosmology; the second type is scattering boundary
conditions, as used in Feynman diagrams, applicable to asymptotically
flat space-time.  The first yields a differential-equation form of the
theory, dual to the integral version appropriate to the second.  Here,
the first (Dirac) approach is found to be extremely streamlined for
the calculation of loop amplitudes in these locally supersymmetric
theories. By contrast, Feynman-diagram methods have led to
calculations which are typically so large as to be
unmanageable. Remarkably, the Riemannian quantum amplitude for
coordinate boundary conditions in $N = 1$ supergravity (without
matter) is exactly semi-classical, being of the form ${\rm exp}(-I/\hbar)$,
where $I$ is the classical action, allowing for the presence of
fermions as well as gravity on the boundaries.  Even when supermatter
is included, typical one-loop amplitudes are often very simple,
sometimes not even involving an infinite sum or integral.
Specifically, the boundary conditions considered for a number of
concrete one-loop examples are set on a pair of concentric 3-spheres
in Euclidean 4-space.  In the non-trivial cases the amplitudes appear
to be exponentially convergent.  

\bigskip
\leftline{\twelverm 1.\quad \uppercase{Two alternative types of
boundary condition}}
\nobreak\smallskip\noindent

 One possibility is to use `coordinate' or `quantum cosmology'
    boundary conditions. For example,    
    every undergraduate first learns quantum mechanics in terms of
    these variables.  He or she is taught about the Schr\"odinger wave
    function $\psi(x,t)$.  For the kind of boundary-value problem
    studied here, the analogue is to specify $x = x_{1}$ at $t =
    t_{1}$ and $x =x _{2}$ at $t = t_{2}$, and to ask for the
    amplitude to go between these data.  A more advanced undergraduate
    or graduate student would learn that this can be computed as a
    Feynman path integral [1], or, alternatively, can be found in
    principle by solving the Schr\"odinger equation given the boundary
    conditions.  Feynman showed that these two dual integral and
    differential formulations are equivalent, in that, for example,
    the path integral obeys the Schr\"odinger equation with the correct
    boundary conditions.  In the celebrated book of Feynman and Hibbs
    [1] many examples of the calculational power of the Feynman path
    integral in ordinary quantum mechanics are given.  At the same
    time, any reader will know that there are many other types of
    problem in quantum mechanics to which the Schr\"odinger approach
    may be much better suited. It should be clear, then, that the
    choice of boundary conditions and method can be purely a pragmatic
    one. Of course, Dirac has taught us that beautiful and elegant
    mathematics often leads to the best physics.  We shall see below
    that this applies to the case of locally supersymmetric theories. 

    It should be pointed out that Dirac and Feynman themselves were
    certainly not insistent on the primacy of one approach over the
    other.  Indeed, one could say that Dirac more or less invented the
    path integral, at least for infinitesimal time separations, as
    described in a paper published remarkably in 1933 in a Russian
    journal (but written in English) [2].  When Feynman learnt about
    this, he thought about it intensively. This led eventually to his
    celebrated Princeton PhD thesis during the war -- the rest is
    history. Conversely, but entirely consistently (since Feynman was
    an ardent admirer of Dirac), the last major project on which
    Feynman worked concerned the ground state of Yang-Mills theory in
    2 + 1 dimensions, treated by Dirac canonical methods [3].
    Feynman attached great weight to this work, and spent at least
    three years on it.  One can conclude that Feynman was not
    slavish about methods, as are some of his followers.

    A further consideration is that our universe is an evolving
    cosmological model, and not asymptotically flat at all.  Hence,
    one cannot even set up an `infinity' region in which to describe
    the familiar scattering problems of particle physics.  Another
    consideration is that  detectors in particle physics experiments
    are at distances of order meters or tens of meters from the
    source, not at infinity.  Therefore,
    it appears that we are forced, for the purposes of comparison with
    experiment or observation, to use cosmological boundary
    conditions. Indeed, the relation between `coordinate' and
    scattering boundary conditions (the second type of familiar
    boundary condition), allowing for gravity, is quite
    problematic.  One might attempt to construct scattering in- and
    out- states in regions of space-time which are not quite at
    infinity, by taking outer products of single-particle
    harmonic-oscillator states of the linearized theory with
    zero-particle states in all the asymptotic directions not occupied
    by ingoing or outgoing particles.  If one considers the
    one-particle wave function for gravitons, one sees that, with a
    very small probability, one may have an arbitrarily large
    gravitational wave excitation in that mode.  If the amplitude of
    the wave is sufficiently large, then the approximate classical
    infilling space-time will be very different from the nearly-flat
    space-time that was originally assumed, and non-linearities will
    totally change the nature of the quantum amplitude.  Clearly, the
    process of taking the limit in which the asymptotic regions are
    taken to infinity is a very awkward one, and needs much further
    detailed investigation.  This may very well account for the
    differences in the divergence structure of quantum amplitudes for
    the same theory, when one  adopts the two different types of boundary
    condition above, which will be seen below.  It is probable that
    such difficulties are much less acute, say, in Yang-Mills theory,
    where such ambiguities have not so far been detected.

\bigskip
\leftline{\twelverm 2.\quad \uppercase{Supergravity and its divergences}}
\nobreak\smallskip\noindent

    In this paper, we shall adopt the Dirac approach to the
    quantization of constrained systems, which Dirac developed
    approximately between 1950 and 1965, particularly with a view to
    the quantization of gravity [4].  Dirac's approach was
    subsequently taken up by Wheeler and DeWitt [5,6], around 1967-8.
    From this work there stemmed a `first era' of quantum cosmology,
    which lasted until around 1975 [7].  Subsequently, in 1983, 
    Hartle and Hawking made their famous proposal for the ground 
    state of the universe, based on the Riemannian Feynman path
    integral approach to quantum gravity [8].  

    When supergravity is treated by Dirac's canonical quantization
    method, one studies physical quantum states such as
    $\Psi(e^{AA'}_{i}(x), \psi^A_i(x))$ [9].  Here, $e^{AA'}_i$ are the
    spatial components $(i = 1,2,3)$ of the tetrad $e^{AA'}_\mu =
    \sigma^{AA'}_a e^a_\mu$, where $\mu = 0,1,2,3$ is a world index
    and $a = 0,1,2,3$ is a tetrad index.  Further, $A = 0,1$ and $A' =
    0',1'$ are two-component spinor indices, and $\sigma^{AA'}_a$ are
    the Infeld-van der Waerden symbols.  The odd Grassmann quantities 
    $\psi^A_\mu$ and $\tilde\psi^{A'}_\mu$ describe the
    four-dimensional gravitino field.  With $\mu$ replaced by $i$, one
    has the fermionic Hamiltonian dynamical data.  In $N = 1$
    supergravity, the Dirac approach requires that a physical state
    should be annihilated by the generators $S^A(x), \bar S^{A'}(x)$
    of local supersymmetry and the generators $J^{AB}(x), J^{A'B'}(x)$
    of local tetrad rotations.  It is essentially trivial to satisfy
    the $J$ constraints, since they simply describe the invariance of
    the wave function under local rotations; equivalently, all
    physical wave functions must be made from spinor Lorentz
    invariants. Explicitly: 

    $$\eqalignno{&S_A = i\hbar ^{3s}D_i\bigl[{\delta\over
    {\delta\psi^A_i}}\bigr] +
    {1\over 2} i\hbar^2\kappa^2{\delta\over{\delta
    {e^{AA'}}_i}}\bigl[{D^{BA'}}_{ji}{\delta\over{{\delta\psi^B}_j}}\bigr] 
    &(2.1)\cr
    &\bar S_{A'} = \epsilon^{ijk}e_{AA'i}~~ ^{3s}D_j{\psi^A}_k + {1\over
    2}
    \hbar\kappa^2 {\psi^A}_i{\delta\over{\delta e^{AA'}_i}.}&(2.2)\cr}$$

    Here, $\kappa^2 = 8\pi$, $^{3s}D_i$ is the torsion-free
    three-dimensional covariant derivative on spinors [9].
    ${D^{AA'}}_{jk} = -2i h^{-{1\over 2}}{e^{AB'}}_k e_{BB'j} n^{BA'}$,
    with $h = det(h_{ij})$, the three-metric $h_{ij}$ being equal to
    the corresponding components $g_{ij}$ of the four-metric.
    Further, the unit outward normal $n^\mu$ corresponds to the spinor
    $n^{BA'} = {e^{BA'}}_\mu n^\mu$. 

    Since there are a number of younger members of the audience who
    did not live through the epoch of supergravity, I should now
    summarize what is known about the ultraviolet divergences of $N =
    1$ supergravity, first without and then with gauged
    supermatter. Let us start with what is known in the case of
    scattering boundary conditions (i.e., by means of Feynman
    diagrams).  We have to examine the historical record as far back
    as the rule of Rameses II; Egyptian mathematics lasted for three
    thousand years and has many achievements.  The relevant record for
    the one-loop case is found on a very crumpled papyrus roll;  of
    course, you have to be able to read hieratics in order to decipher
    it.  But, there is no doubt it reads that pure supergravity is
    finite at one loop.  For two loops, we have to move forward into
    the Middle Ages, where in a monastery of the twelfth century there
    was found a frayed parchment document in Mediaeval Latin in which a
    mathematically-minded monk discovered that pure supergravity was
    also finite at this level.  Beyond two loops, we look in vain at
    the historical record, whether in manuscript or stored in some
    computer. As far as anyone can tell, the question as to whether
    pure supergravity has a divergence at three loops has never been
    resolved.  This limitation to our knowledge appears to be the
    result simply of human frailty.  Perhaps if someone lived to the
    age of Methuselah, they might stand a chance of managing the
    enormous Feynman-diagram calculation.  [For those of you who do
    not come from the Judaeo-Christian-Islamic tradition, Methuselah
    was the oldest man in the Bible; he lived to 969 years of age.]
    The notion that supergravity at three loops is divergent is no
    more than a myth.

    For pure supergravity with quantum cosmology boundary conditions,
    one finds no ultraviolet divergences at any order.  This can be
    seen from a general argument [9] and its workings can be examined
    in more detail in examples such as those at the end of this
    paper.  

    Turning to the more general case of $N = 1$ supergravity with
    gauged supermatter, as described clearly in the second edition of
    the book of Wess and Bagger [10], this is always divergent at all
    loops in the scattering formulation [11].  But an argument similar
    to that mentioned above shows that all amplitudes using quantum
    cosmology boundary conditions are finite, as in pure
    supergravity.  This is partly based on the property that, when the
    boundary data including supermatter are purely bosonic, the
    amplitude is exactly semi-classical:  $\Psi = {\rm exp}(-I/\hbar)$.  

    This connects with a very general pure-mathematical problem.  To
    take the simplest example, consider the `Hartle-Hawking' classical
    boundary-value problem for pure Einstein gravity.  Suppose one
    takes a boundary manifold of topology $S^3$ with a given
    Riemannian three-metric $h_{ij}$ and an interior region with the
    usual topology.  One then asks whether there is a unique (up to
    diffeomorphism) Riemannian four-metric $g_{\mu\nu}$ on the
    interior, agreeing with the boundary metric and obeying the vacuum
    Einstein equations $R_{\mu\nu} = 0$.  It is easy to prove this in
    the case of small perturbations of the round sphere, using a
    fixed-point method or equivalently the implicit function theorem [12],
    by first studying the problem linearized about flat Euclidean
    space. Some related results, found by working close to known
    manifolds, have recently been established [13,14].
    Of course, the resulting metrics are the analogue of weak
    gravitational waves which are perturbations of Minkowski
    space-time.  On the other hand, there has to date been no general
    study of this problem for large deformations of the sphere.  One
    would think that this would be a wonderful arena for pure
    mathematicians!  (G. Gibbons has pointed out that, if one replaces
    the usual interior topology by a suitable bundle topology, then
    for sufficiently deformed boundary data there may be two, not one,
    classical solutions inside, of Taub-Bolt type [15].)  
    It is interesting to compare the historical
    situation with regard to the hyperbolic Cauchy evolution problem
    for the Einstein equations in the usual Lorentzian case, a
    question which Einstein himself might well have asked.  By 1953,
    Yvonne Choquet-Bruhat had already begun to attack this problem
    [16].
    Its resolution had reached an advanced state by 1970 [17].
    By comparison, Riemannian quantum gravity began with the work
    of Hartle and Hawking on the positive-definite Schwarzschild
    solution in 1975 [18], and the boundary-value problem above, associated
    with the same authors, dates from 1983 [8].

\bigskip
\leftline{\twelverm 3.\quad \uppercase{Suitable boundary data}} 
\nobreak\smallskip\noindent

    Starting with the Riemannian classical boundary-value problem for
    pure Einstein gravity,  one might at first think that the simplest
    boundary conditions would be to specify the three-metric on two
    nearly-planar three-surfaces at different imaginary time
    coordinates, measured at spatial infinity, assuming asymptotic
    flatness at spatial infinity.  Unfortunately, for weak
    perturbations of flat Euclidean four-space, the resulting
    four-metric $g_{\mu\nu}$ does not depend in a very smooth way on
    the boundary data $h_{ij}$ in a neighbourhood in which $h_{ij}$ is
    close to the flat metric $\delta_{ij}$, as pointed out by Stephen
    Hawking. To see this, suppose that the lower boundary is
    intrinsically flat, and that the upper boundary has a curved
    intrinsic three-metric, such that it can be embedded in flat
    Euclidean four-space with the other boundary as the lower
    boundary.  Then note that one could have equally well turned the
    upper boundary surface upside down, and still had a flat solution
    of the same boundary-value problem.  A more detailed investigation
    shows that the small deviations in the classical four-metric go roughly
    as the square root of the deviations in the boundary three-metric,
    in this neighbourhood. Therefore, it is not really practicable to
    use Fourier analysis in studying this perturbation problem.  

    Instead, one takes the next-simplest possibility:  two concentric
    three-spheres, such that, in the unperturbed configuration, the
    inner sphere has radius $\alpha$ and the outer sphere has radius
    $\beta$.  It is then appropriate to decompose all perturbations in
    terms of harmonics on $S^3$.  The original treatment, for bosonic
    perturbations, was given by Lifschitz in 1946.  Greater
    detail is given by Lifschitz and Khalatnikov [19].  Workers in
    cosmology will be familiar with these:  for spin $s = 0$, one has
    density perturbations, for $s = 1$, rotational perturbations, and
    for $s = 2$, one has cosmological gravitational waves.  For
    example, one can write the scalar modes as 
    $$\eqalign{Q^n_{lm} &= \Pi^n_l(\chi)Y_{lm}(\theta,\phi)\cr
    {Q_{|i}}^{|i} &= -(n^2-1)Q,\cr}\eqno(3.1)$$
    where $\chi,\theta,\phi$ are standard coordinates on the unit
    three-sphere as defined in Eq.(3.3), and $\Pi^n_l$ obeys a
    suitable radial equation.  

    To cover harmonics of all spins $s = 0, {1\over 2}, 1,{3\over 2}, 2,$ it is
    best to use two-component spinors.  As a preliminary, we need to
    evaluate integrals of powers of $x^a$ over the unit three-sphere,
    where $a$ is a tetrad index.  Suppose we are given $n = 2m (m =
    0,1,...)$.  Define 
    $$C_n = \int d\Omega u^n,\eqno (3.2)$$
    where $d\Omega $ is the measure on the unit three-sphere, and
    where 
    $$\eqalign{x &= \sin \chi  \sin\theta \cos \phi ,\cr
    y &= \sin\chi \sin\theta \sin\phi, \cr
    z &= \sin\chi \cos\theta,\cr
    u &= \cos\chi.}\eqno(3.3)$$
    One finds, using [20], 
    $$C_n = 4\pi^2 {{1 \times 3...\times (n-3)\times (n-1) \times
    1}\over {2 \times 4...\times (n-2)\times n\times
    (n+2)}}.\eqno(3.4)$$
    As $n\rightarrow $, one finds from Stirling's formula [20] that 
    $$C_{n = 2m}\sim{{2\pi^{3\over 2}}\over{(m +
    1)^{1\over 2}}}.\eqno(3.5)$$
    Now consider 
    $$\int d\Omega x^{a_1}x^{a_2}...x^{a^{2n}} = D_n
    \delta^{(a_1a_2}
    \delta^{a_3a_4}...\delta^{a_{2n-1}a_{2n}}),\eqno(3.6)$$
    with
    $$\int d\Omega u^{2n} = D_n =C_n.\eqno (3.7)$$
    We shall need the spinor version of these equations, giving 
    $$\eqalign{& \int d\Omega x^{A_1A'_1}x^{A_2A'_2}...x^{A_{2m}A'_{2m}} =
    {C_{2m}\over{(2m)!}}\bigl(\epsilon ^{A_1A_2}
    \epsilon^{A_3A_4}...\epsilon ^{A'_1A'_2}\epsilon^{A'_3A'_4}...\cr
    & + \hbox{\rm all  permutations  on  both  primed  and  unprimed  
    indices.}\bigr)\cr}\eqno(3.8)$$
    
    The tensor and spinor harmonics on $S^3$ can now be described in a
    uniform way:   
    (a) $s = 0$. Instead of the Lifschitz-Khalatnikov description
    above, one can write a normalized harmonic $\phi^{npq}$ on the
    unit sphere as 
    $$\phi^{npq} =
    T_{(A_1...A_n)(A'_1...A'_n)}x^{A_1A'_1}...x^{A_nA'_n},\eqno(3.9)$$
    where $T_{...}$ is a constant array of the form 
    $$\eqalign{&T_{...} = E_n\bigl(T_{00...011...10'0'...0'1'1'...1'}\cr
    &+ \hbox{\rm the 
    remainder of the}\quad (n!)^2\quad \hbox{\rm permutations  on
    both  primed  
    and  unprimed  
    indices}\bigr).\cr}\eqno(3.10)$$
    Here, there are $p$ zeros and $q$ primed zeros, and 
    the quantity $T_{...}$ on the righthand side of Eq.(3.10) is
    numerically equal to $1$.  The normalization constant $E_n$ is
    fixed by the requirement 
    $$\int d\Omega\; \phi^{npq}\bar\phi^{npq} = 1 =
    C_{2n}2^n(n!)^2|E_{np}|^2.\eqno(3.11)$$
    One can check from this definition that $\phi^{npq}$ obeys the
    harmonic equation Eq.(3.1).  
    
    (b) $s = {1\over 2}.$  These harmonics are described in more detail in
    [21].  There are normalized positive frequency harmonics 
    $$\rho_A^{npq} =  _{(A}T_{A_1...A_n)(A'_1...A'_n)}
    x^{A_1A'_1}...x^{A_nA'_n}\eqno(3.12)$$
    similar to the scalar harmonics in (a) above.  One can check that
    they obey the eigenvalue equation
    $$e^{AA'j}~~ ^{3s}D_j\rho _A = -(n + {3\over 2})_en^{AA'}\rho_{A},
    \eqno(3.13)$$
    where 
    $$_en^{AA'} = -in^{AA'}\eqno(3.14)$$
    is the Euclidean normal.  Similarly, there are positive frequency
    primed harmonics 
    $$\sigma_{A'}^{npq} = _{A'}T_{A_1...A_n
    A'_1...A'_n}x^{A_1A'_1}...x^{A_nA'_n}, \eqno(3.15)$$
    where $_{A'}T_{...}$ is again totally symmetric on primed and
    unprimed indices.  The negative frequency modes are of the form 
    $$\tau_A \propto _e  {n_A}^{A'} \sigma_{A'}, \quad \mu_{A'}\propto
    _e{n^A}_{A'} \rho_A,\eqno(3.16)$$
    which for example obey
    $$e^{AA'j}\;^{3s}D_j\tau_A = + (n + {3\over 2})
    _en^{AA'}\tau_A.\eqno(3.17)$$ 
    
    (c) $s = 1$.  The harmonics are of the form 
    $$v_{AA'}^{npq} = _{AA'}T_{A_1...A_nA'_1...A'_n} x^{A_1A'_1}...x
    ^{A_nA'_n}, \eqno (3.18)$$
    where, as always $T$ is totally symmetric.  

    (d)  $s = {3\over 2}$.  The true gravitino data are given by the
    harmonics 
    $$\rho_{(ABC)}^{npq} = _{ABC}T_{A_1...A_nA'_1...A'_n}
    x^{A_1A'_1}...x^{A_nA'_n}, \eqno(3.19)$$ 
    with the usual symmetry.  

   (e)  $s = 2$.  The true graviton data are analogously given by 
   $$e_{AA'BB'}^{npq} = _{AA'BB`}T_{A_1...A_nA'_1...A'_n}
   x^{A_1A'_1}...x^{A_nA'_n}, \eqno(3.20)$$
   again with total symmetry.

\bigskip
\leftline{\twelverm 4.\quad \uppercase{Loop amplitudes in $N = 1$
 supergravity}}
\nobreak\smallskip\noindent 

   As remarked earlier, if we only had gravitational perturbations on
   the two spherical boundaries, then the full amplitude in quantum $N
   = 1$ supergravity would be exactly semi-classical, of the form 
   $exp (-I/\hbar)$, where $I$ is the classical gravitational action.
   For any hope of non-trivial quantum effects, one should put
   fermionic data on the boundaries.  With our boundary data set on
   the concentric pair of spheres, the simplest fermionic weak-field
   case is to specify a harmonic, ${\tilde\psi}_{A'B'C'}^{MPQ}$ on the
   inner boundary, and the corresponding harmonic ${\psi_{ABC}}^{MPQ}$
   on the outer boundary.  Of course, one could simulate scattering by
   taking a linear combination of two harmonics on each boundary, but
   the qualitative conclusions will not be greatly changed.  

   One proceeds by applying the supersymmetry constraint $S_C \Psi =
   0$ to the amplitude
   $$\Psi\sim(A + \hbar A_1 +\hbar^2 A_2 + ...){\rm exp}
   (-I_{class}/\hbar).\eqno(4.1)$$
   At the lowest order $\hbar^0$, one obtains the classical constraint
   $$S_C = 0,\eqno (4.2)$$
   which is automatically satisfied by virtue of the classical field
   equations.  At the next order, $\hbar^1$ one finds 
   $$\eqalign{&\quad^{3s}D_i\bigl[{{\delta(log A)}\over
   {\delta\psi^C_i(x)}}\bigr]\cr
   & + {1\over 2}\hbar\kappa^2 {{\delta}\over{ \delta
   e^{AA'}_i(x)}}\bigl[D^{BA'}_{ji} {{\delta
   I}\over{\delta\psi^B_j(x)}}\bigr] = 0.\cr}\eqno(4.3)$$    
   Here, $\kappa^2 = 8\pi.$  The right-hand object in square brackets
   is in fact the classical
   $\tilde\psi^{A'}_i(x)$, as one finds by consideration of the 
   canonical fermionic momentum [9].  Note the double functional
   derivative at the same point $x$, but with respect to one fermionic
   and one bosonic argument.  It turns out that this does not lead to
   the kind of infinities which are inevitably present in the
   canonical quantization of pure Einstein gravity, through terms of
   the kind  
   $$...+ {{\delta ^2 \Psi}\over {\delta h_{ij}(x)\delta h_{kl}(x)}}...,$$ 
   as in the following example.  

   The classical solutions are derived from the Euclidean action of $N
   = 1$ supergravity; 
   $$\eqalign{I &= \int_{{\rm
   VOL}}d^4x\bigl[{{-1}\over{2\kappa^2}}({\rm det}\; e)R
   \cr
   &+
   {1\over
   2}\epsilon^{\mu\nu\rho\sigma}(\tilde\psi^{A'}_{\mu}e_{AA'\nu}D_{\rho} 
   \psi^A_{\sigma}  + h.c.)\bigr]\cr
   &+ \int_{{\rm BDRY}}d^3x\bigl[{{-1}\over{\kappa^2}}h^{1\over 2}(tr K) +
   \epsilon^{ijk}\psi^A_i
   e_{AA'j}\tilde\psi^{A'}_k\bigr].\cr}\eqno(4.4)$$
   Here, $K_{ij}$ is the second fundamental form [9] and $tr K =
   h^{ij}K_{ij}$.  Also, at a classical solution the volume integral
   vanishes, and the action $I$ resides only in the boundary
   integral.  

   For the perturbation problem involving $\psi_{ABC}^{MPQ}$ etc.,
   there are several contributions to $I$ of the type $e\tilde\psi
   \psi$, needed for the right hand side of Eq.(4.3).  Let us take a
   typical one, of the largest possible size, arising from $K_{ij}$,
   contributing to the covariant derivative $D_{\rho}$ in the volume
   integral.  One can calculate the change in $\tilde\psi_{A'B'C'}$,
   due to the addition of (say) a small graviton perturbation $\delta
   e_{AA'BB'} (x')$ by integrating the Rarita-Schwinger equation
   radially. The general graviton perturbation on the outer surface
   can be written as 
   $$\delta  e_{AA'BB'}(x') = \Sigma_{NRS}\;c_{NRS}\;
   e_{AA'BB'}^{NRS}(x').\eqno(4.5)$$ 
   Then, by orthogonality, 
   $$c_{NRS} = \int d\Omega\; \delta
   e_{AA'BB'}(x')\;e_{CC'DD'}^{NRS}(x')\;n^{AC'}n^{BD'}n^{CA'}n^{DB'}.
   \eqno(4.6)$$
   Hence,
   $${{\delta c_{NRS}}\over{\delta e_{AA'BB'(x)}}} =
   e_{...}^{NRS}(x)nnnn.\eqno(4.7)$$
   where the indices on the right hand side are straightforward to
   calculate.  The above-mentioned change in $\tilde\psi_{A'B'C'}$
   depends on all the constants $c_{NRS}$.  The total change of course
   involves a radial as well as an angular integral.  But it turns out
   that the resulting dependence on the boundary radii $\alpha$ and
   $\beta$ is unimportant. 

   One finds that 
   $$\eqalign{& \tilde\psi^{A'}_i(x) \sim \Sigma_{NRS}\int d\Omega'c_{NRS}
   e_{...}^{(NRS)}(x')\tilde\psi_{...}^{MPQ}(x')\cr
   & \times \hbox{\rm terms
   proportional to the normal}\quad n,\cr
   & \hbox{\rm the alternating symbol}\quad
   \epsilon \quad\hbox{\rm
   and the
   background spatial tetrad}\quad e_{...}^{i}.\cr}\eqno(4.8)$$
   Hence, 
   $$\eqalign{&{\delta\over{\delta
   e^{AA'}_i(x)}}[\tilde\psi^{A'}_i(x)]\cr
   &\sim \Sigma_{NRS}\int d\Omega'\;e_{...}^{NRS}(x')
   e_{...}^{NRS}(x)\tilde\psi_{...}^{MPQ}(x')\cr
   & \times \hbox{\rm  other  terms,  as 
   above}.\cr}\eqno(4.9)$$
   But, by completeness, the sum over $NRS$ reduces to a product of
   delta functions with respect to the indices.  Since on the left
   hand side there is only one free index $A$, the right hand side
   must consist of a multiple of $\tilde\psi_{ABC}(x)$ with the two
   indices $BC$ contracted, namely zero.  Hence the corresponding
   ${\rm log}
   A$ is also zero.  A similar result would have been obtained if we
   had allowed for a linear combination of two harmonics on the
   boundaries.  

   A more general argument, leading to a similar conclusion, can be
   given (say) when the boundary data on both spheres consist only of
   a weak-field mixture of spin-3/2 harmonics.  This depends on
   examining the left hand term of Eq.(4.3), instead of the right hand
   term.  This arises because one cannot make ${\rm log} A$ out of some
   quantity contracted with $\psi_{ABC}^{MPQ}$, etc., because as one
   can check, 
   $$^{3s}D_i\psi_A^{MPQi} = 0,\eqno(4.10)$$
   where 
   $$\psi_A^{MPQi} = e^{BB'i}{n^C}_{B'}\psi_{ABC}^{MPQ}.\eqno(4.11)$$
   This is analogous to the property [19]
   $${S^{i}}_{|i} =  0, \eqno(4.12)$$
   where $S^i$ represents a generic spin-1 harmonic in the language of
   Lifschitz and Khalatnikov  [19].   Hence one must have ${\rm log} A = 0$
   in this case.  Of course, one could instead have checked that the
   right hand term in Eq.(4.3) was also zero.  

   One could only make a non-zero ${\rm log} A$ if the boundary data
   contained a spin-1/2 part $\psi_A^{MPQ}$, which is forbidden in our
   simple example by the $\tilde S_{A'} = 0 $ classical constraint:
   $$\tilde S_{A'} = \epsilon^{ijk}e_{AA'i} ^{3s}D_j\psi^A_k + {1\over
   2}  
   i\kappa^2\psi^A_i  p_{AA'}^i = 0,\eqno(4.13)$$
   where $p_{AA'}^i$ is the momentum conjugate to $e^{AA'}_i$ [9].
   The next simplest generalization of these boundary data would be to
   include in addition a weak-field graviton mode $e_{AA'BB'}^{RST}$
   on the outer boundary (say).  The classical constraint (4.13) will
   enforce an extra non-zero spin-1/2 term on the outer boundary,
   which is at least quadratic in the gravitino and graviton
   perturbations (starting with a cross term).  Again, this cannot
   match the right hand term in Eq.(4.3).  The same holds for more
   general gravitino and graviton perturbations.  We conclude that 
   $$\Psi = {\rm exp} (-I_{\rm class}/\hbar)\eqno(4.14)$$
   for pure $N = 1$ supergravity.  

   This might, at first sight, seem shocking.  It says that there are
   no quantum corrections for $N =1$ supergravity with these boundary
   conditions.  All the dynamics resides in the classical motion.
   However, it was previously known [9] that Eq.(4.14) held for
   purely bosonic boundary data, and so it does not seem unreasonable
   that the same should be true when one includes fermionic data,
   given the local supersymmetry of the theory.  Further, all our
   experimental knowledge of loop effects comes, of course, from
   particle physics at `low' energies, which only involves
   non-gravitational interactions.

\bigskip
\leftline{\twelverm 5.\quad \uppercase{$N = 1$ supergravity with
   gauged supermatter}}
\nobreak\smallskip\noindent

   Due to the work of many authors, not listed here, the general
   locally  supersymmetric $N = 1$ model of gravity interacting with a
   gauge theory has been found [10].  Because of the huge amount of
   local symmetry -- local supersymmetry, local coordinate invariance,
   local tetrad rotation invariance, and local gauge invariance --
   these models are of a very restricted type.  The Lagrangian 
   $\cal L$ can be split as 
   $${\cal L} = {\cal L}_{kin} + {\cal L}_{pot}, \eqno(5.1)$$
   where ${\cal L}_{kin}$ is determined once the symmetry group such
   as SU(2), SU(3), etc., and the gauge coupling constant $g$ are
   specified.  The remaining part ${\cal L}_{pot}$ depends on a
   potential $P$ which is a function of the scalar fields.  For
   simplicity, in this section we shall set $P = 0$.  

   In the simplest non-trivial case, with SU(2) gauge group [10], there
   is one complex scalar $(a, a^*)$ with K\"ahler potential
   $$K = log (1 +a a^*)\eqno (5.2)$$
   and K\"ahler metric 
   $$g_{11^*} = {{\partial^2K}\over{\partial a\partial a^*}} =
   {{1}\over{(1 + aa^*)^2}}.\eqno(5.3)$$
   Now, 
   $$ds^2 = {{da da^*}\over{(1 + aa^*)^2}}\eqno(5.4)$$
   is the metric on a unit two-sphere (really, $CP^1$).  The point
   with $a = a^* = 0$ may be regarded as the North pole, and the point
   $a = a^* = \infty$ is then the South pole; there is nothing
   preferred about these points -- for example, the scalar field may
   move freely through $a = \infty$.  The connection between the
   K\"ahler scalar part of the theory and the gauge theory is that
   the isometry group $SU(2)$ for the scalars is, then, the gauge group
   of the full theory.  

   The other fields in the kinetic theory may be summarised as
   follows.  There is a spinor field $(\chi _A, \tilde\chi_{A'})$,
   which has no Yang-Mills index, and which is the partner of
   $(a,a^*)$.  The Yang-Mills field $v_{\mu}^{(a)}$, with $(a) =
   1,2,3$ has fermionic partners $(\lambda _A^{(a)},
   \tilde\lambda_{A'}^{(a)}).$  As usual, one also has the tetrad
   $e_{AA'\mu}$ and the gravitino $(\psi_{A\mu},\tilde\psi_{A'\mu}).$
   The relevant Lagrangian may be found in Wess and Bagger [10].  

   If, say, one wanted to extend this model to $SU(3)$, one could use
   the corresponding K\"ahler metric given in [22].  For $SU(n)$,
   one can similarly use [15].  

   As in the case above of pure $N = 1$ supergravity, one proceeds to
   find loop terms iteratively using the quantum supersymmetry
   constraint $S_A \Psi = 0.$  In the present $SU(2)$ case, the
   operator $S_A$ has the general structure 
   $$\eqalign{&\quad {\rm const.}
   ^{3s}D_i\bigl({{\delta}\over{\delta\psi^C_i(x)}}\bigr)\cr
   & + {\rm
   const.}{{\delta}\over {\delta
   e^{AA'}_i(x)}}\bigl[D^{BA'}_{ji}{{\delta}\over{\delta\psi^B_j(x)}}\bigr]\cr
   & +{\rm const.}\epsilon^{ijk}e_{AB'k}
   n^{BB'}F_{ij}^{(a)}\bigl({{\delta}\over{\delta\lambda^{(a)B}}}\bigr)\cr
   &+ {\rm const.} n^{BB'}
   e_{AB'i}{{\delta}\over{{\delta \; v^{(a)}}_i}}\bigl({{\delta}\over
   {\delta\lambda^{(a)B}}}\bigr)\cr
   & + {\rm const.}g_{11^*}(\tilde{\cal D}_i
   a^*){e_A}^{B`i}\bigl({{\delta}\over
   {\delta\tilde\chi^{B'}}}\bigr)\cr
   & + {\rm const.} {n_A}^{A'}{{\delta}\over{\delta
   a}}\bigl({{\delta}\over{\delta \tilde\chi^{A'}}}\bigr)\cr
   & + {\rm higher-order\quad terms}.\cr}\eqno(5.5)$$
   Here, the fermionic coordinates are being regarded as $\psi^A_i,
   \lambda^{(a)B}, \tilde\chi^{A'},$ while $F_{ij}^{(a)}$ are the
   spatial components of the Yang-Mills field strength.  The covariant
   derivative $\tilde{\cal D}_i a^*$ is defined in [10].  

   Once a loop term has been found (iteratively, if necessary), one 
   must further check that the conjugate quantum constraint $\bar
   S_{A'}\Psi = 0$ is also satisfied.  This occurs in the examples
   below. 

   The Dirac approach to the computation of loop terms in such a
   locally supersymetric theory is extremely streamlined by comparison
   with the corresponding path-integral calculation.  As can be seen
   from Eq.(5.5), in the Dirac approach one only needs to concentrate
   on related fermionic and bosonic partners to find the dependence of
   the amplitude on those variables.  This removes many
   complications.  In contrast, a path-integral treatment would
   inevitably involve integration over the relatively large number of
   fields, and one would always have to be verifying cancellation
   effects between bosons and fermions.  

   Now consider some examples of loop calculations, with the usual
   pair of spherical boundaries, in the simplest $SU(2)$ model.  This
   work has been carried out jointly with M.M. Akbar.  Note, from
   [10],
   that there is a negative cosmological constant, of order $g^2$
   in the theory.  Strictly, this implies that the background
   four-geometry is a Riemannian space of constant negative curvature.
   In this case, the angular harmonics used are the same, but the
   radial dependence of the corresponding classical solutions for a
   given principal quantum number $n$ changes from a power law to an
   exponential.  This makes no qualitative difference to the outcomes
   of the calculations below.  Alternatively, the reader may wish to
   imagine that $g$ is exceedingly small.  

   Example (1)
   Find the one-loop correction in the case that the data on the inner
   sphere, of radius $\alpha$, are a harmonic of the first kind of
   spinor field above:  $\chi_A = \chi_A^{PMN},$ while the data on the
   outer sphere are the corresponding $\tilde\chi_{A'} =
   \tilde\chi_{A'}^{PMN}.$  From Eq.(5.5), we see that we need to make
   a variation $\delta a^*(x)$, which can be expanded in scalar
   harmonics as 
   $$\delta a^*(x) = \Sigma_{QST}\; c_{QST}\;\bar\phi^{QST}.\eqno(5.6)$$
   Ignoring the details caused by the radial dependence (which lead to
   a possible dependence on $\alpha$ and $\beta$, one finds
   schematically from Eq.(5.5) that 
   $$\eqalign{&(\partial_i \phi^{RWX}(x))e_{BA'}^i n
   ^{BB'}\bigl({{\delta {\rm
   log}A}\over{\delta\tilde\chi_{B'}(x)}}\bigr)\cr
   & = \Sigma_{YZL}\; n_{CA'}\; \phi^{B(YZL)}(x)e_{BB'}^i (\partial
   _i\phi^{RWX})\bar \phi^{B'(YZL)}\phi^{C(PMN)}(x).\cr}\eqno(5.7)$$
   Now note that, when one fixes $Y$ in the summation, but sums over
   all $ZL$ consistent with this, the term $\phi^{B(YZL)}e_{BB'}^i\bar
   \phi^{B'(YZL)}$ cancels out.  Hence, in this case, ${\rm log}A =
   0.$  

   This example may look too simple, in that we have only taken one
   harmonic for both surfaces.  However, the same conclusion arises
   when one considers a `scattering' problem with two harmonics added
   together on each surface.  

   Example (2)  
   The intention here is to illustrate a gravitational effect on a
   one-loop term $A.$  The data chosen are as in Example (1), except
   that one adds in a weak field scalar harmonic $a(x) =
   \phi^{RWX}(x)$ on the outer surface.  Clearly, the presence of $a$
   will induce a non-trivial gravitational field at quadratic and
   higher orders in $a$, which will then contribute to $I, A,...$
   Since the calculation is a little complicated, we take here the
   simplest non-trivial case with the lowest spinor harmonic $P = 0$,
   giving $\chi_A = {\rm constant}$ if we were in flat Euclidean
   four-space.  

   One finds, without detailed attention to the radial dependence, 
   $$\eqalign{&{{\delta({\rm log}A)}\over{\delta\tilde\chi_{B'}(x)}}\cr
   &= \Sigma_{QNY}\phi^{QNY}(x)\phi^{(Q+R,N+W,Y+X)}(x)\chi^{B}(x) n_{BB'}\cr
   &\times \bigl[{{QR}\over{(Q+R)(Q+R+1)}}\bigr]{{(Q+R)! Q!
   R!}\over{C_{Q+R}C_QC_R C_{2(Q+R)}2^{Q+R}(2Q+2R)!}}.\cr}\eqno(5.8)$$
   When one does include the radial dependence, it only makes a
   difference of $O(1)$, except that, as usual in quantum gravity, a
   single power of $\hbar$ is associated with two negative powers of
   radius.  This means that the present one-loop term is smaller than
   the kind of one-loop term which might be found by studying the
   interactions between particles of spins $0, {1\over 2}, 1$ by a factor
   of order $({\rm Planck\; length/}\beta)^2.$  If, say, $\beta$
   were 1 cm., our factor would be down by ${10}^{-66}$ on a typical
   one-loop factor.  

   To understand the rate of convergence of the sum in Eq.(5.8), one
   uses Stirling's formula [20], which shows that, for large $Q$, the
   sum has the form 
   $$\Sigma_Q ({\rm slow}) 2^{-{\rm const.}Q}.\eqno(5.9)$$
   Here, the `slow' terms are typically polynomial, and one sees that
   the convergence is exponential;  this is of course much faster than
   in any Feynman diagram.  One might similarly ask about the
   corresponding two- and higher-loop terms.  Because of the way in
   which the spinor indices combine in the spinor harmonics above, the
   dominant structure is always the same:  the sum of the terms inside
   the factorial signs on the top line is always the same as the sum
   of the corresponding terms on the bottom line.  But the terms on
   the bottom line are always combined in larger fragments.  Because
   of the way in which Stirling's theorem works, this means that one
   always will find negative exponentials for large values of the
   principal quantum numbers, which will overwhelm any `slow'
   polynomial terms arising from `gravitational vertices'.  It is of
   course the dreaded polynomials in the momentum which lead to the  
   non-renormalizability of Einstein quantum gravity.  

   Had we, in Example (1), say, taken data which give a non-trivial
   sum on the right hand side, for ${\rm log} A$ or for higher loops,
   but which do not perhaps involve gravitational interactions in the
   classical action, we would still have found the same dominant
   structure in the sum on the right hand side, leading again to an
   exponential convergence.  

   Example (3) 
   Here we choose the quark-like fermionic data, given by a harmonic 
   $\tilde\lambda_{(PMN)}^{(a)B'}$ on the inner sphere, and $\lambda
   _{(PMN)}^{(a)B}$ on the outer sphere.  Recall that the bosonic
   partner of $\lambda^{(a)B}$ is $v_{m}^{(b)}.$  The constraint, 
   as given by Eq.(5.5), yields
   $$\eqalign{& \epsilon^{ijk} e_{BA'k}F_{ij}^{(a)}\bigl({{\delta{\rm
   log} A}\over{\delta\lambda^{(a)B}(x)}}\bigr)n^{BB'}\cr
   &\sim\Sigma_{QNP}...\epsilon^{abc}v_l^{(a)(QNP)}v_l^{(b)(QNP)}...\cr
   & = 0.\cr}\eqno(5.10)$$
   Hence, ${\rm log} A  = 0$ in this case also.  

   Of course, there are many interactions between particles of spins
   $0, {1\over 2}, 1$, which one would expect to lead to various loop
   effects.  However, the `gravitational' example (2) should be
   sufficient to illustrate what happens in such cases.  

   Since the loop behaviour of this model appears reasonable, it seems
   worthwhile to investigate the model further with regard to its
   physical consequences, and to try to predict effects which are
   observable at accelerator energies.

\bigskip
\leftline {\uppercase {Acknowledgements}}
\nobreak\smallskip\noindent
  
   I would like to acknowledge valuable discussions with Gary Gibbons
   and Mohammad Akbar.  I am also very grateful to Kathleen Wheeler
    for help in preparation of the manuscript.
   Finally, Norma Sanchez is owed warm thanks for her most efficient
   but discrete organization of the INTAS meeting in May, 1998 at
   which this paper was delivered.

\bigskip
\centerline{\bigbf REFERENCES}
\nobreak\smallskip\noindent

\hang\noindent [1]  Feynman, R.P. and Hibbs, A.R. (1965) {\it
   Quantum Mechanics and Path Integrals} (McGraw Hill, New York).

\hang\noindent [2]  Dirac, P.A.M. (1933) The Lagrangian in quantum
   mechanics. Phys. Z. der Sowjetunion, Band 3, Heft 1, 64-72,
   reprinted in {\it Quantum Electrodynamics}, ed. Schwinger,
   J. (Dover, New York, 1958). 
   
\hang\noindent [3]  Feynman, R.P. (1981) The qualitative behavior of
   Yang-Mills theory in 2+1 dimensions. Nucl. Phys. B {\bf 188}, 479-512.

\hang\noindent [4]  Dirac, P.A.M.  (1965) {\it Lectures on Quantum 
   Mechanics}(Academic Press, New York).

\hang\noindent [5]  Wheeler, J.A.  (1968) Superspace and the nature of quantum
   geometrodynamics, in {\it Battelle Rencontres 1967},
   ed. C.M. DeWitt and J.A. Wheeler (Benjamin, New York).

\hang\noindent [6]  DeWitt, B.S. (1967) Quantum theory of gravity
   1. The canonical 
   theory. Phys. Rev. {\bf 160}, 1113-48.

\hang\noindent [7]  Ryan, M.P. and Shepley, L.C. (1975)  {\it Homogeneous
   Relativistic Cosmologies}  (Princeton University Press, Princeton).
 
\hang\noindent [8]  Hartle, J.B. and Hawking, S.W. (1983) Wave function of the
   universe.  Phys. Rev. {\bf D28}, 2960-75.

\hang\noindent [9]  D'Eath, P.D. (1996) {\it Supersymmetric Quantum
   Cosmology} (Cambridge University Press, Cambridge).

\hang\noindent [10]  Wess, J. and Bagger, J. (1992) {\it Supersymmetry and
   Supergravity} (Princeton University Press, Princeton).

\hang\noindent [11]  van Nieuwenhuizen, P. (1981) Supergravity.
   Phys. Rep. {\bf  68}, 189-398.
 
\hang\noindent [12]  Choquet-Bruhat, Y., DeWitt-Morette, C.M., and
   Dillard-Bleick, 
   M. (1982) {\it Analysis, Manifolds and Physics} (North Holland, 
   Amsterdam).

\hang\noindent [13]  Schlenker, J.-M. (1998) Einstein manifolds with convex
   boundaries.  Orsay preprint. 

\hang\noindent [14]  Rivin, I. and Schlenker, J.-M. (1998)  The  Schl\"afli
   formula and Einstein manifolds. IHES preprint.

\hang\noindent [15]  Eguchi, T., Gilkey, P.B. and Hanson, A.J. (1980)
   Gravitation, 
   gauge theories and differential geometry, Phys. Rep. {\bf 66},
   213-393.  

\hang\noindent [16]  Bruhat, Y. (1962) The Cauchy problem.  In {\it
   Gravitation: 
   An Introduction to Current Research}, ed. Witten, L. (Wiley, New
   York).

\hang\noindent [17]  Hawking, S.W. and Ellis, G.F.R. (1973) {\it The
   Large Scale 
   Structure of Space-Time}  (Cambridge University Press, Cambridge).

\hang\noindent [18]  Hartle, J.B. and Hawking, S.W. (1976)  A path-integral
   derivation of black hole radiance.  Phys. Rev. {\bf D13},
   2188-2207.
  
\hang\noindent [19]  Lifschitz, E.M. and Khalatnikov, I.M. (1963)
   Investigations in 
   relativistic cosmology.  In Advances in Physics {\bf 12}, 185-249.

\hang\noindent [20]  Abramowitz, N. and Stegun, A.I.  (1965)  {\it Handbook of
  Mathematical Functions} (Dover, New York).

\hang\noindent [21]  D'Eath, P.D. and Halliwell, J.J. (1987) Fermions
   in quantum 
   cosmology. Phys. Rev. {\bf D35}, 1100-23.

\hang\noindent [22]  Gibbons, G.W. and Pope, C.N. (1978) ${\rm CP}^2$ as a
   gravitational instanton.  Commun. Math. Phys. {\bf  61}, 239-48. 

\end